\documentclass[prd,aps,nofootinbib,onecolumn,showpacs,10pt]{revtex4}
\usepackage{graphicx,epsfig}

\usepackage{epsf}
\usepackage{graphicx}

\begin{document}

\title{  \bf  $ B_c B_c J/\psi $ Vertex Form Factor at Finite Temperature in the Framework of QCD Sum Rules Approach}
\author{E. Yazici$^{*1}$, H. Sundu$^{*2}$, E. Veli Veliev$^{*3}$ \\
$^{*}$Department of Physics, Kocaeli University, 41380 Izmit,
Turkey\\
$^1$e-mail:enis.yazici@kocaeli.edu.tr
$^2$email:hayriye.sundu@kocaeli.edu.tr
\\$^3$ e-mail:elsen@kocaeli.edu.tr\\}

\begin{abstract}

The strong form factor of the $B_{c} B_{c}J/\Psi$ vertex is calculated in the framework of the QCD sum rules method at finite temperature. Taking into account additional operators appearing at finite temperature, thermal Wilson expansion is obtained and QCD sum rules are derived. While increasing temperature, the strong form factor remains unchanged up to $T\simeq100~MeV$ but slightly increases  after this point. After $T\simeq160~MeV$, the form factor suddenly decreases up to $T\simeq170~MeV$. The obtained result of the coupling constant by fitting the form factor at $Q^2=-m^2_{offshell}$ at $T=0$ is in a very good agreement with the QCD sum rules calculations at vacuum. Our prediction can be checked in the future experiments.

\end{abstract}

\maketitle

\section{Introduction}
The three-meson vertex are important quantities in the phenomenological theories of hadron physics. Knowledge of the mesons' coupling constants or form factors plays important role to understand mesons' interactions, meson decays, structure of the exotic mesons and so on. The form factors depend on the momentum as $Q^2$ and some input parameters. In the last years meson form factors $D^*D\pi$\cite{Nav}, $D^*D\rho$\cite{2}, $D^*DJ/\psi$\cite{3}, $D^*_sD_sJ/\psi$\cite{4}, $B^*_sBK$\cite{5}, $B^*_sB_s\eta^{(')}$\cite{6}, $B^*_cB_c\Upsilon$\cite{7}, $B^*_cB_cJ/\psi$\cite{8} and other vertices widely discussed in the literature (for detailed information see \cite{8} and \cite{9}). Moreover, many new heavy meson states have been observed and investigated in the recent years. The investigation of strong interaction of heavy hadrons among themselves has received increasing attention and this situation motivates us to investigate bottom mesons interactions.

On the other hand, beginning from the CERN-SPS data analyses of heavy-ion collisions, Quark Gluon Plasma (QGP) phase has taken many physicists' attention. Coming to this day, new experimental results pointing a QGP phase in p-Pb collisions at the LHC have been announced \cite{10}. The understanding of thermal meson properties before phase transition requires to study the temperature dependencies of meson form factors. The knowledge on temperature dependence of form factors is very important for interpretation of heavy-ion collision experiments and understanding QCD vacuum. For instance, comparing to its value in vacuum, we see that the value of the form factor decreases suddenly around the deconfinement temperature. We consider this as a sign of a possible phase transition. The aim of extending sum rules to finite temperature is to understand thermal properties of hadrons. Because understanding thermal properties of QCD vacuum allows us to clarify properties of observable universe. Moreover, theoretical studies on QCD at finite temperature is necessary to explain the experimental results obtained from the heavy ion collisions. It is assumed that the increase of density and temperature of media where the hadrons are formed rearrange physical properties of hadrons such as decay widths and masses. Also, it is known that heavy mesons have different behaviours when temperature of the medium increases (see \cite{tempp} and references therein). These motivated us to investigate thermal properties of mesons.

Form factors at vertices of three heavy mesons at low energies are not be able to calculate using perturbative approaches. The QCD sum rules method is one of the most effective tools considering non-perturbative nature of hadronic properties \cite{11}, since it does not have any model-dependencies. The extending of QCD sum rules method to finite temperature has been made in the paper \cite{12}. This extension based on two basic assumptions, that the Operator Product Expansion (OPE) and notion of quark-hadron duality remains valid at finite temperature, but the vacuum condensates must be replaced by their thermal expectation values \cite{13,14,15,16}. The thermal QCD sum rules approach has been extensively used for studying masses and decay constants of both light and heavy mesons as a reliable and well-established method \cite{13,14,15,16,17,18,19,20,21,22,23,24,25,26}. But, in the literature, there are few theoretical works devoted to the analysis of the hadron vertex form factors at finite temperature.

The aim of this paper is to investigate strong form factor $B_{c} B_{c}J/\psi$ vertex in the framework of the QCD sum rules method at finite temperature. The key concepts of the strong form factor calculations in sum rules are three-point correlation functions. In order to calculate form factors, one must have the temperature dependencies of masses and decay constants of the hadrons in the vertex. Since we aim to extend strong form factors to finite temperatures, in this study, we choose mesons whose masses and leptonic decay constants have been investigated in thermal QCD sum rules approach \cite{25,26}.

The outline of this paper is as follows: in the  next section, the QCD sum rules for the form factors of the related mesons are obtained in the framework of the  QCD sum rules at finite temperature. In Section III, our numerical predictions for the strong coupling constant and comparison of the results with the existing predictions of the other studies are presented.

\section{QCD Sum Rules for the Strong Form Factors}
In this section, we obtain the sum rules for the strong form factor of $B_{c} B_{c} J/\Psi$ vertex in the QCD sum rules framework at finite temperature. We
 evaluate the temperature-dependent version of the three-point correlation function,
\begin{eqnarray}\label{correl.func.1}
\Pi_{\mu}(q,T) =i^2\int d^{4}x d^{4}y e^{ip' \cdot x}e^{iq \cdot y}{\langle} {\cal T}[J^{B_{c}} (x) J_{\mu}^{J/\psi } (y) J^{B_{c} \dagger}(0)]{\rangle},
\end{eqnarray}
where $\cal T$ is the time ordering product in the correlation function and $T$ is temperature,
$J^{B_{c}}(x)=\overline{c}(x)\gamma_{5}b(x)$ and $J_{\mu}^{J/\psi}(x)=\overline{c}(x)\gamma_{\mu}c(x)$ are the interpolating currents of the
related mesons. The thermal average of any operator $A$ is expressed as:
\begin{eqnarray}\label{A}
{\langle}A {\rangle}=\frac{Tr(e^{-\beta H} A)}{Tr e^{-\beta H}},
\end{eqnarray}
where $H$ is the QCD Hamiltonian and $\beta=1/T$ is the inverse of temperature $T$.

The correlation function is calculated in two different ways: first, in the hadronic language which is called as physical side; and second, by using the QCD terms which is called OPE side. After equating these two different representations of correlation functions, we acquire the QCD sum rules for the strong form factors. In order to suppress the higher state contributions and continuum contributions we apply a double Borel transformation with respect to the variables $p^2$ and $p'^2$ to physical side and OPE side.

In the phenomenological side, a complete set of intermediate  states with the same quantum numbers with current $J(x)$ are inserted between the currents in Eq.
(\ref{correl.func.1}) and the integral is performed.  As a result, we obtain
\begin{eqnarray}\label{phen1}
\Pi_{\mu}^{phen}(q^2,0)=\frac{\langle 0\mid J^{B_{c}}(0) \mid B_{c}(p')\rangle \langle 0\mid J^{J/\psi}_\mu(q) \mid J/\psi(q)\rangle \langle B_{c}(p') J/\psi(q) \mid B_{c}(p)\rangle \langle B_{c}(p) \mid J^{B_{c}}(0)\mid
 0\rangle}{(q^2-m^{2}_{J/\psi}) (p^2-m^{2}_{B_{c}}) (p'^2-m^{2}_{B_{c}})}
&+& \cdots,
\end{eqnarray}
 where ($\cdots$) represents the  higher states and continuum contributions. Creation of the pseudoscalar and vector mesons from the vacuum can be presented in terms of the leptonic decay constants of mesons, $f_{M}$:
\begin{eqnarray}\label{lep1}
\langle 0\mid J^{B_{c}}(0) \mid B_{c}(p')\rangle=\frac{f_{B_{c}} m^{2}_{B_{c}}} {m_{b}+m_{c}},
\end{eqnarray}

\begin{eqnarray}\label{lep2}
\langle 0\mid J^{J/\psi}_\mu(q) \mid J/\psi(q)\rangle=f_{J/\psi} m_{J/\psi} \epsilon_{\mu} ,
\end{eqnarray}

\begin{eqnarray}\label{lep3}
\langle B_{c}(p') J/\psi(q) \mid B_{c}(p)\rangle=g_{B_{c}B_{c}J/\psi}(p'-q)\cdot \epsilon ,
\end{eqnarray}
where, $m_{M}$ and $f_{M}$ are the mass and decay constant of the related meson, respectively. Note that Eqs. (3)-(6) are valid also at finite temperature, hence, the final representation for the physical side can be written in terms of the temperature-dependent mass and decay constants as:

\begin{eqnarray}\label{phen1}
\Pi_{\mu}^{phen}(q^{2},T)=\frac{g_{B_{c}B_{c}J/\psi}(q^{2},T) m^{4}_{B_{c}}(T) m_{J/\psi}(T) f_{B_{c}}^2(T) f_{J/\psi}(T)}{(m_{b}(T)+m_{c}(T))(q^2-m^{2}_{J/\psi}(T)) (p^2-m^{2}_{B_{c}}(T)) (p'^2-m^{2}_{B_{c}}(T))} \left[-2p'_{\mu}+\left(\frac{m^2_{B_{c}}(T)-q^2}{m^2_{J/\psi}(T)}\right)p_{\mu}\right]+\cdots,
\end{eqnarray}
where $f_M(T)$ and $m_M(T)$ are temperature-dependent parameters.

\begin{figure}[h!]
\begin{center}
\includegraphics[width=5cm]{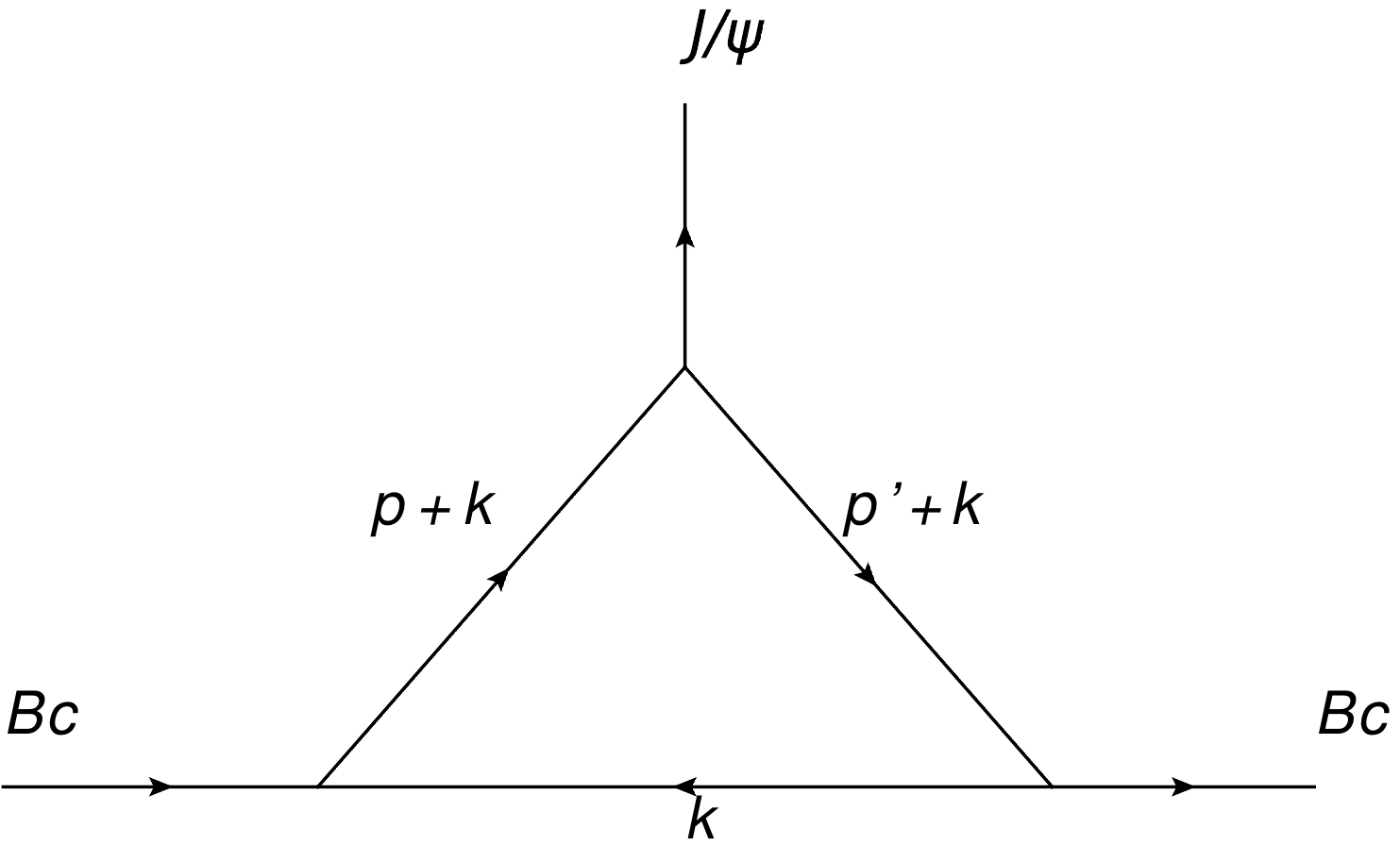}
\end{center}
\caption{Bare loop diagram.}
\label{fig1}

\begin{center}
\includegraphics[width=5cm]{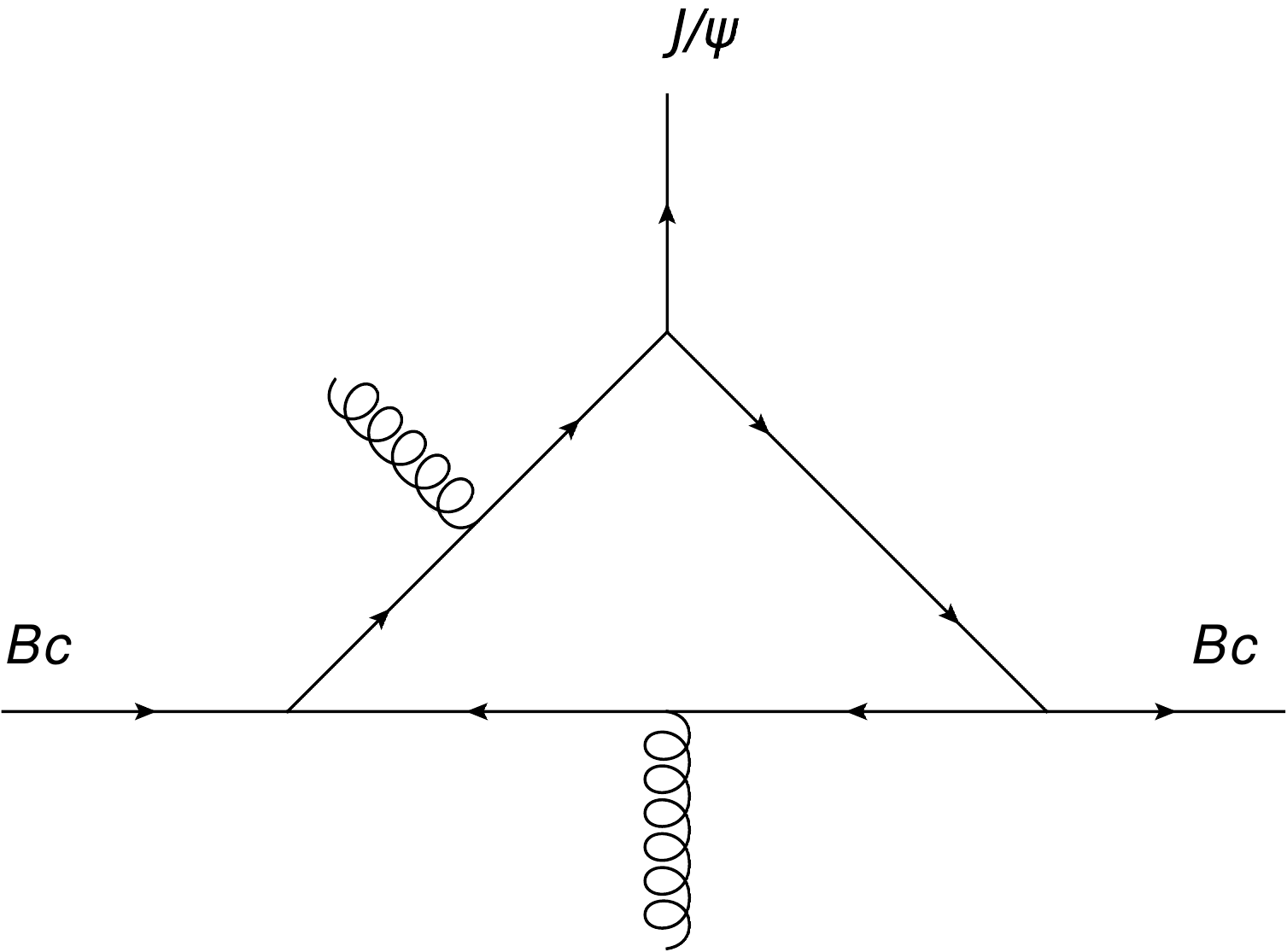}
\includegraphics[width=5cm]{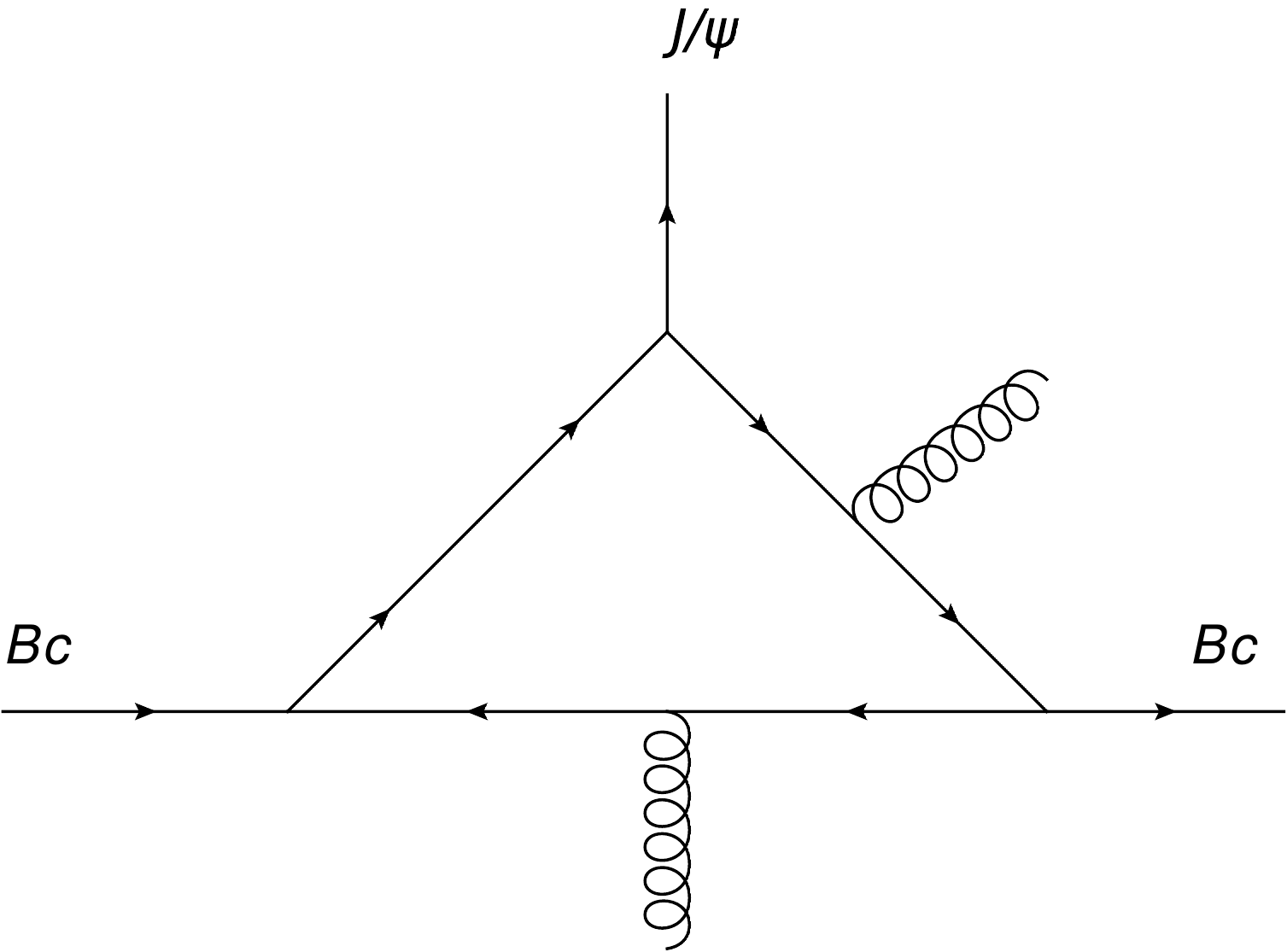}
\includegraphics[width=5cm]{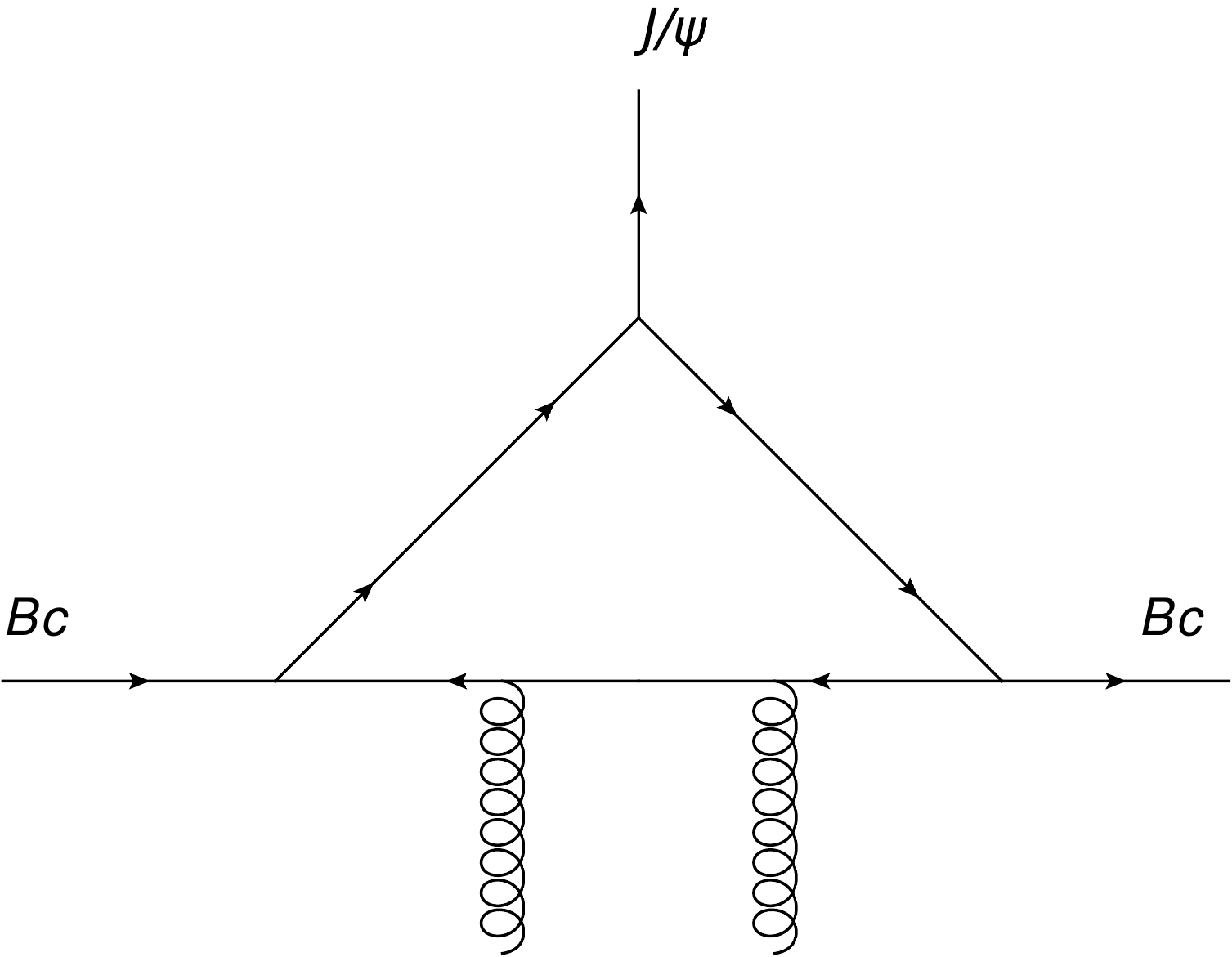}
\includegraphics[width=5cm]{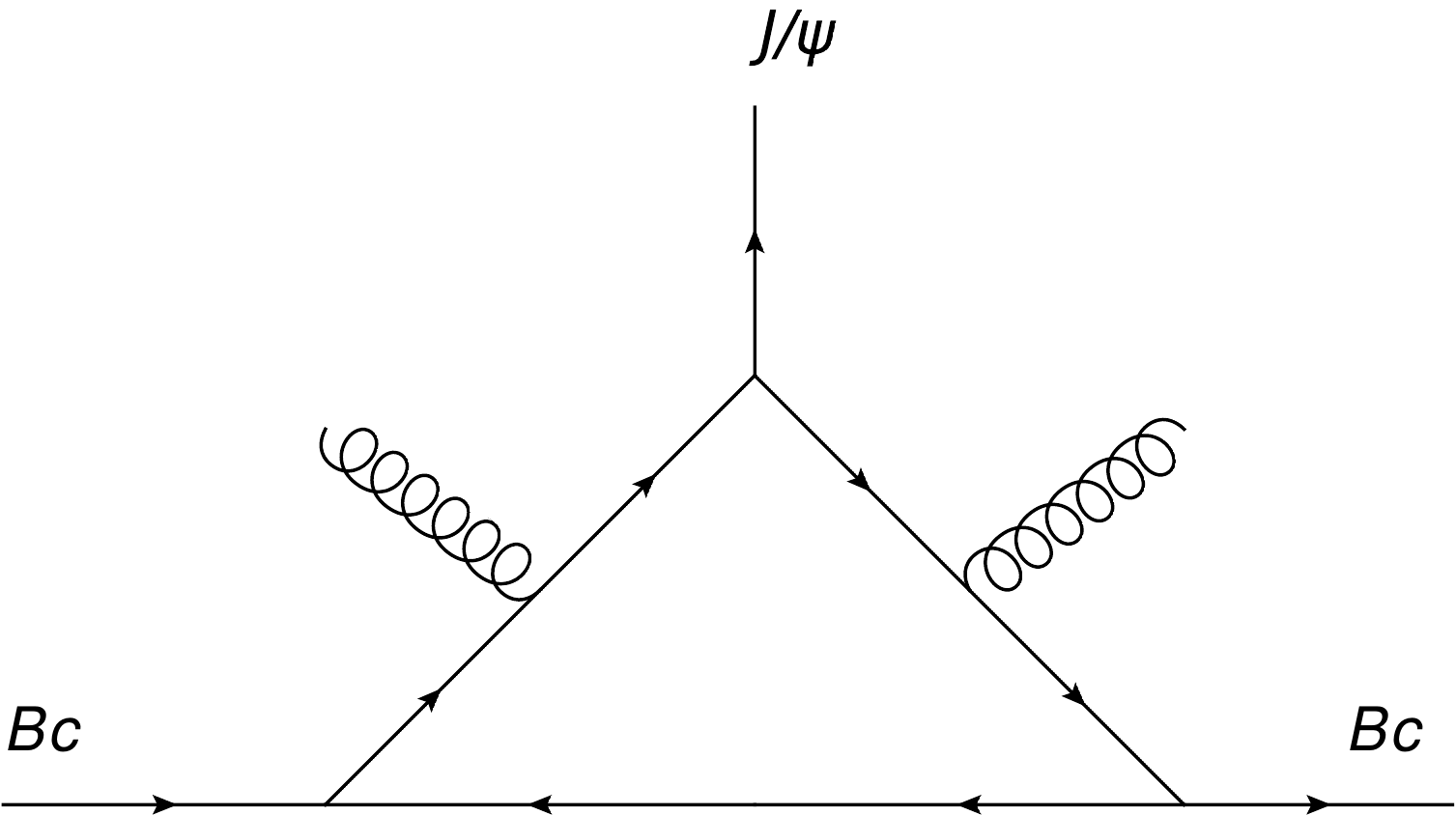}
\includegraphics[width=5cm]{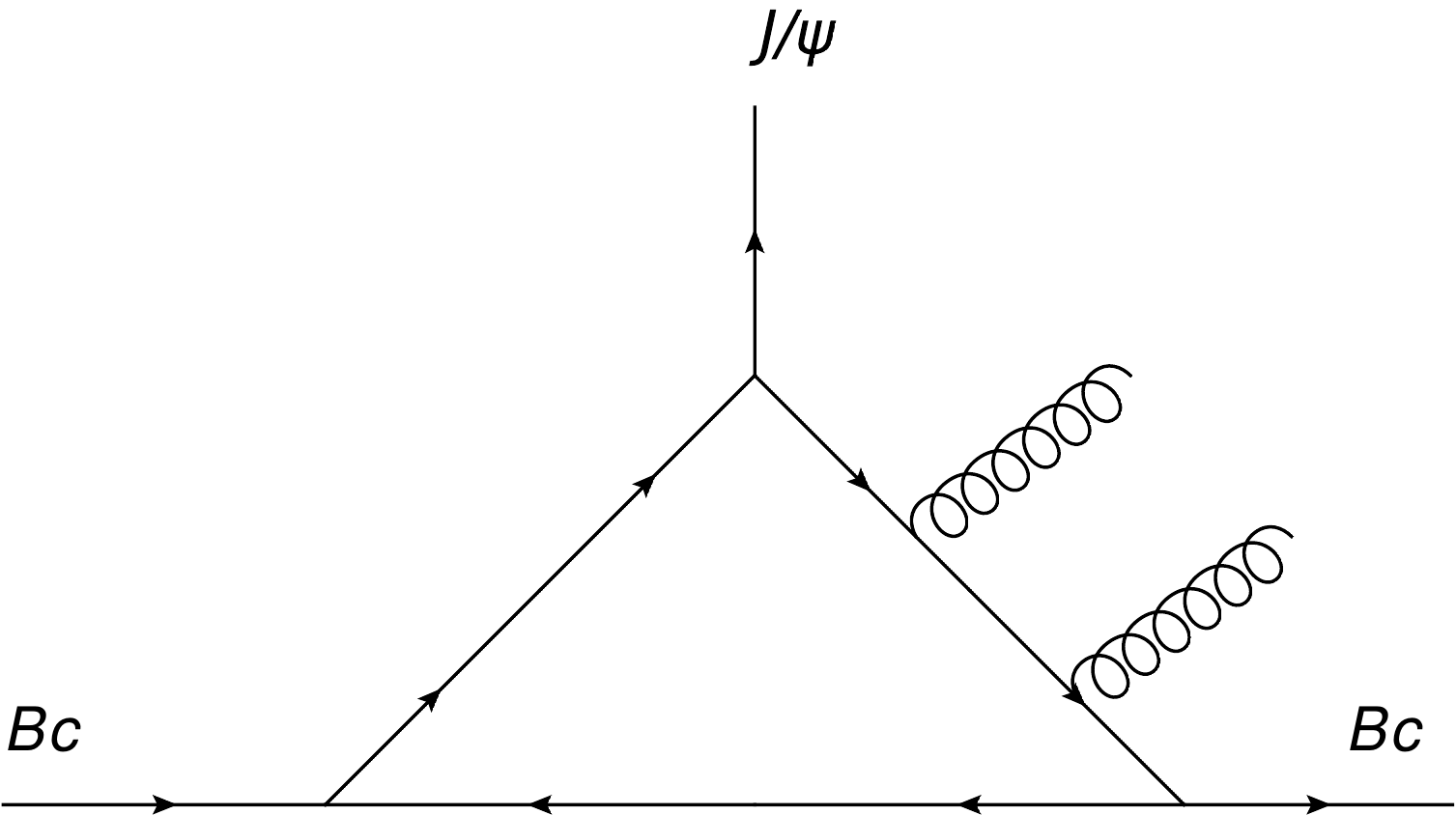}
\includegraphics[width=5cm]{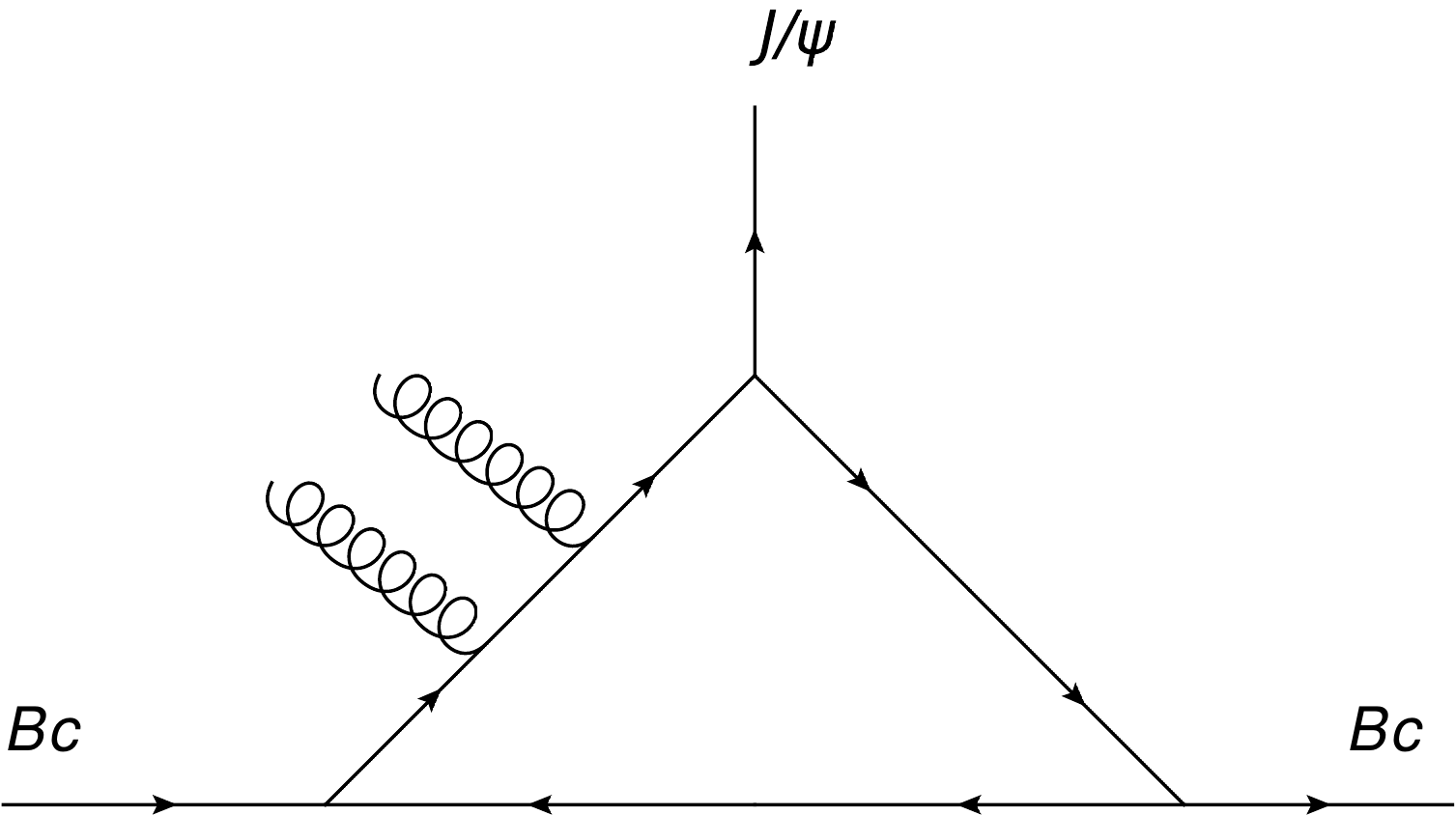}
\end{center}
\caption{Diagrams corresponding to two-gluon condensates.}
\label{fig2}
\end{figure}

 In the QCD side, the correlation function is calculated in deep Euclidean region, $q^2\ll-\Lambda^{2}_{QCD}$, by using operator product expansion (OPE) where the short and long distance effects are separated:
\begin{eqnarray}\label{correl.func.QCD1}
\Pi_{\mu}^{QCD}(q^2,T) =\Pi_{\mu}^{pert}(q^2,T)+\Pi_{\mu}^{nonpert}(q^2,T).
\end{eqnarray}
The short distance contributions are  calculated by using the perturbation theory, contrary to the long distance effects which are represented in terms of the vacuum expectation values of some operators having different mass dimensions.  Moreover, we chose $p_{\mu}$ as the structure and write the terms which are proportional to $p_{\mu}$. Concluding, by using dispersion integral representation, the equation above can be written as:
\begin{eqnarray}\label{correl.func.QCD1}
\Pi^{QCD}(q^2,T) =\int ds \int ds' \frac{\rho(s,s',q^2,T)}{(s-p^2)(s'-p'^2)}+\Pi^{nonpert}(q^2,T),
\end{eqnarray}
where $\rho(s,s',q^2,T)$ is called the spectral density. To obtain the spectral density, the Feynman diagram in figure
(\ref{fig1}) which is called the bare loop diagram is used.
 After straightforward calculations, spectral density is obtained as:
\begin{eqnarray}\label{rhoper}
\rho(s,s',q^2,T)&=&\frac{-N_{c}}{\sqrt{(s+s'-q^{2})-4ss'}} \frac{1}{-q^{4}+(s-s')^{2}-q^{2}(s+s')}\nonumber\\
&\times & [ 2m_{c}^{3}(m_{b}-m_{c})s'+(m_{b}^{4}-m_{b}^{3}m_{c})(q^{2}-s+s')+ss'(-2m_{c}^{2}+q^{2}-s+s')\nonumber\\
&+& m_{b}^{2}(q^{2}-s-s')(-m_{c}^{2}+q^{2}-s+s')+m_{c}^{2}s'(-2m_{b}^{2}+2m_{b}m_{c}-q^{2}+s+s')\\
&-& m_{c}(m_{b}-m_{c})((q^{2}+s-s')s'+m_{b}^{3}(q^{2}-s+s')+m_{c}^{2}(-q^{2}+s+s')\nonumber\\
&+& m_{b}m_{c} (-q^{4}+m_{c}^{2}(-q^{2}-s-s')+s(-s+s')+q^{2}(2s+s') ) ) ]\nonumber.
\end{eqnarray}

In non-perturbative part, the main  contribution comes from the two
gluon condensates since the heavy quark condensates can be safely neglected. Because they are suppressed by inverse powers of the quark masses.
The two-gluon condensate diagrams are shown in
figure (\ref{fig2}). While calculating non-perturbative
contributions, Fock-Schwinger gauge $x^{\mu}A^{a}_{\mu}(x)=0$ is used. Vacuum gluon field in momentum space and quark-gluon-quark vertices used in calculations are given as:
\begin{eqnarray}\label{Amu}
A^{a}_{\mu}(k)=-\frac{i}{2}(2 \pi)^4 G^{a}_{\rho
\mu}(0)\frac{\partial} {\partial k_{\rho}}\delta^{(4)}(k),
\end{eqnarray}
\begin{eqnarray}\label{qgqver}
\Gamma=ig \frac{\lambda^{a}}{2}\gamma_{\mu}A^{a}_{\mu}(k),
\end{eqnarray}
where $k$ is the gluon momentum.

  In order to calculate non-perturbative contributions, ${\langle}Tr^{c}G_{\alpha \beta}G_{\mu \nu}{\rangle}$ must be known. By using Lorentz invariance at finite temperature, this expectation value is expressed as:

\begin{eqnarray}\label{qgqver}
{\langle} Tr^{c}G_{\alpha \beta}G_{\mu \nu} {\rangle}&=&\frac{1}{24}\left(g_{\alpha\mu}g_{\beta\nu}-g_{\alpha\nu}g_{\beta\mu}\right){\langle} G^{a}_{\lambda \sigma}G^{a\lambda\sigma} {\rangle}\nonumber\\
&+&\frac{1}{6}\left[g_{\alpha\mu}g_{\beta\nu}-g_{\alpha\nu}g_{\beta\mu}-2(u_{\alpha}u_{\mu}g_{\beta\nu}-u_{\alpha}u_{\nu}g_{\beta\mu}-u_{\beta}u_{\mu}g_{\alpha\nu}+u_{\beta}u_{\nu}g_{\alpha\mu})\right]{\langle} u^{\lambda}\Theta_{\lambda\sigma}^{g}u^{\sigma} {\rangle}.
\end{eqnarray}

 The Borel transformed phenomenological part of the correlation function is obtained as:

\begin{eqnarray}\label{Gammaphys}
\hat{B}\Pi^{phen}(q^2,T)&=& \frac{g_{B_{c}B_{c}J/\psi}(q,T) m_{B_c}^{4}(T) m_{J/\psi}(T) f_{B_c}^{2}(T) f_{J/\psi}(T)}{(m_{b}(T) +m_{c}(T))(q^2-m_{J/\psi}^{2} (T))}\nonumber\\
&\times & \left(\frac{m_{B_c}^{2}(T)-q^2}{m_{J/\psi}^{2}(T)}\right)exp\left[\frac{-m_{B_c}^{2}(T)}{M^2}\right]exp\left[\frac{-m_{B_c}^{2}(T)}{M'^2}\right].
\end{eqnarray}
The non-perturbative part of QCD side in Borel transformed scheme is given by:

\begin{eqnarray}\label{Gammanonper}
\hat{B}\Pi^{nonpert}(q,T)&=& \int_{0}^{1} dx \int_{0}^{1-x} dy \langle \alpha_{s} G^2\rangle F(m_q,M^2,M'^2,T),
\end{eqnarray}
where $F$ function is dependent of quark masses, Borel parameters and also temperature.
After required calculations of the gluon condensate contributions, choosing coefficients which are proportional to $p_{\mu}$ and matching the QCD and the phenomenological sides of the correlation function, we obtain the following sum rule
 for the strong form factor as a function of $Q^2=-q^2$ and temperature $T$.
\begin{eqnarray}\label{coupsum}
g_{B_{c}B_{c}J/\psi}(Q^2,T)=\left\{ \int_{s=(m_{c}+m_{b})^2}^{s_{0}(T)} ds ~ \int_{s'=(m_{c}+m_{b})^2}^{s'_{0}(T)} ds'   ~\rho(s,s',Q^2,T)~exp\left(-\frac{s}{M^{2}}\right)+~exp\left(-\frac{s}{M'^{2}}\right)\hat{B}\Pi^{nonpert}\right\}\nonumber\\
\times \frac{m^2_{J/\psi}(T) (m_{c}+m_{b})^2 (-Q^2-m^2_{J/\psi}(T))}{m^4_{B_{c}}(T)m_{J/\psi}(T)(m^2_{B_{c}}(T)+Q^2)f^2_{B_{c}}(T)f_{J/\psi}(T)}exp\left(\frac{m^2_{B_{c}}(T)}{M^2}\right)exp\left(\frac{m^2_{B_c}(T)}{M'^2}\right),
\end{eqnarray}
where $M^2$ and $M'^2$ are the Borel mass parameters, and $s_{0}$ and $s'_{0}$ are continuum thresholds. The temperature-dependent expressions of continuum thresholds are obtained as \cite{Cheng,Miller}:

\begin{eqnarray}\label{continuum1}
s_{0}(T)=s_{0}\left[ 1- \left( \frac{T}{T_{c}^*}\right)^{8} \right]+(m_{c}+m_{b})^2\left( \frac{T}{T_{c}^*}\right)^{8}~ ,
\end{eqnarray}

\begin{eqnarray}\label{continuum2}
s'_{0}(T)=s'_{0}\left[ 1- \left( \frac{T}{T_{c}^*}\right)^{8} \right]+(m_{c}+m_{b})^2\left( \frac{T}{T_{c}^*}\right)^{8}~ ,
\end{eqnarray}
where critical temperature value is given as $T_c=0.176GeV$ and $T_c^*=1.1 \times T_c$ \cite{Cheng,Miller}.

\section{Numerical analysis}

This section presents the numerical analysis of  the sum rules for the  form factor of the heavy mesons based on thermal considerations. In the analysis, we use the following input parameters,
$m_c=(1.28\pm0.05)GeV$, $m_b=(4.18\pm0.1)GeV$ \cite{pdg} and
${\langle}0\mid\alpha_s G^2  \mid 0 {\rangle}=0.038GeV^4$
\cite{11}. In the calculations, thermal versions of some parameters such as the continuum thresholds and the vacuum condensates are used. Temperature-dependent masses and decay constants are obtained in \cite{25,26} by using thermal QCD sum rules approach. The values of masses at $T=0$ are calculated as $m_{J/\psi}=(3.05\pm0.08)GeV$ and $m_{B_c}=(6.37\pm0.05)GeV$ which are in a very good agreement with experimental data. The temperature-dependent expressions of masses and decay constants are:

\begin{eqnarray}\label{massTemp}
m_{B_{c}[J/\psi]}(T)=A+B ~ exp\left[\frac{T}{C}\right],
\end{eqnarray}

\begin{eqnarray}\label{decayTemp}
f_{B_{c}[J/\psi]}(T)=D+E~ exp\left[\frac{T}{F}\right].
\end{eqnarray}

\begin{table}[h]
\renewcommand{\arraystretch}{1.5}
\addtolength{\arraycolsep}{3pt}
$$
\begin{array}{|c|c|c|c|c|c|c|}
\hline \hline
  ~ & A(GeV) &  B(GeV) & C(GeV) & D(GeV) & E(GeV)& F(GeV) \\
\hline
  B_c &  6.9662    &  -2.8752\times 10^{-5}   &  0.0184  & 0.5047 & -11.4930\times 10^{-5} & 0.0217 \\
 \hline
 J/\psi  &  3.05430  &  -2.4332\times 10^{-5} & 0.0178  &  0.4842  & 9.2392\times 10^{-5}  &  0.2149 \\
  \hline \hline
\end{array}
$$
\caption{Coefficients of the temperature-dependent mass and decay constant functions \cite{25,26}. } \label{tab:mass}
\renewcommand{\arraystretch}{1}
\addtolength{\arraycolsep}{-1.0pt}
\end{table}

 The sum rules expression for the form factor contains four other parameters which are
continuum thresholds $s_0$ and $s'_0$, and Borel masses $M^2$ and $M'^2$.  According to the QCD sum
rules standard, the physical quantities should be independent of Borel  parameters. Hence, the Borel mass intervals should be as stable as possible. The working region for the Borel mass parameters are determined  requiring that not only the higher state and continuum contributions are suppressed, but also the contribution of the highest order operator should be small. As a result, the working region for the Borel parameters are to be $ 18~ GeV^2 \leq M^{(')2} \leq 22~ GeV^2 $. The continuum thresholds are related to the energy of the first exited states of the mesons with the same quantum numbers of the mesons in the vertex. Continuum thresholds can depend on the Borel mass parameters \cite{Lucha}. We choose the intervals $s_0=s'_0=(41-43)~GeV^2$ for the continuum thresholds.

In the calculation we use the rest frame. In this frame,
$u^{\alpha}$ is defined as $u^{\alpha}=(1,~0,~0,~0)$. $\Theta^{g}$
is gluonic part of energy density and described in \cite{20,Cheng,Miller}. We use the lattice results which can be summarised as follows:
\begin{eqnarray}\label{tetag}
\langle \Theta^{g}_{00}\rangle=T^4~ exp\left[113,87T^2-12,2T\right]-10,14T^5(GeV^4).
\end{eqnarray}
Here, temperature unit is GeV and  $0.1\leq T \leq 0.197~GeV$. In \cite{Cheng,Miller}, $\langle G^2\rangle$ is parameterised as:
\begin{eqnarray}\label{G2TLattice}
\langle G^2\rangle=\langle
0|G^2|0\rangle\left[1-1.65\left(\frac{T}{T_{c}}\right)^{8.735}+0.05\left(\frac{T}{T_{c}}\right)^{0.72}\right].
\end{eqnarray}

\begin{figure}[h!]
\begin{center}
\includegraphics[width=10cm]{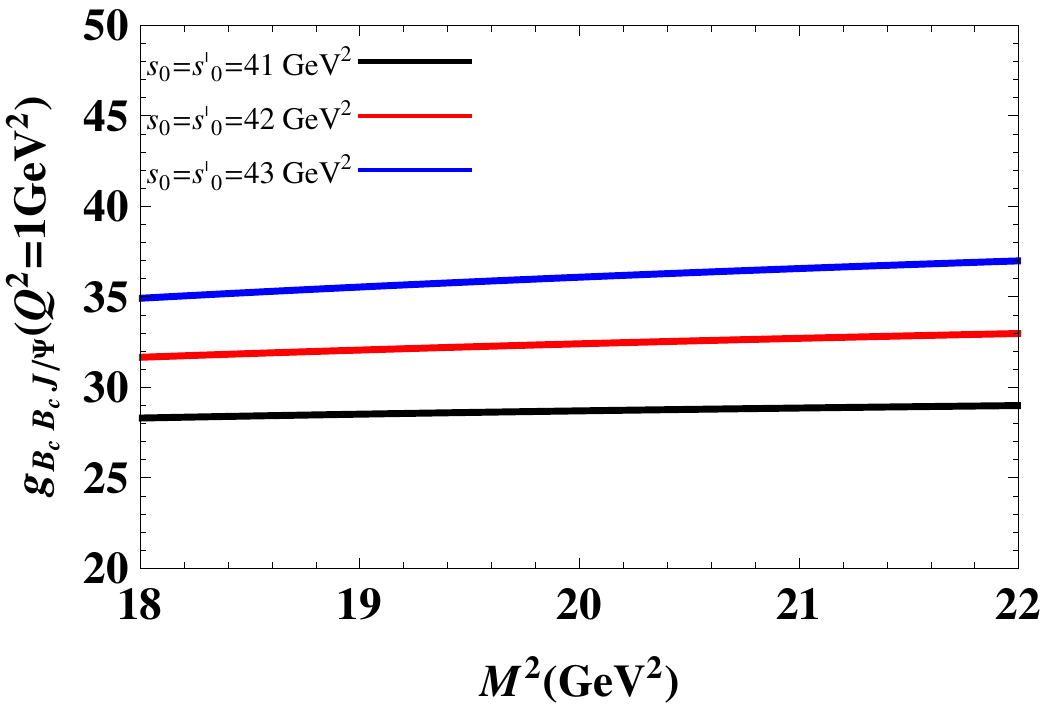}\\
\end{center}
\caption{The dependence of the form factor on $M^{2}$.}
\label{fig3a}
\end{figure}

\begin{figure}[h!]
\begin{center}
\includegraphics[width=10cm]{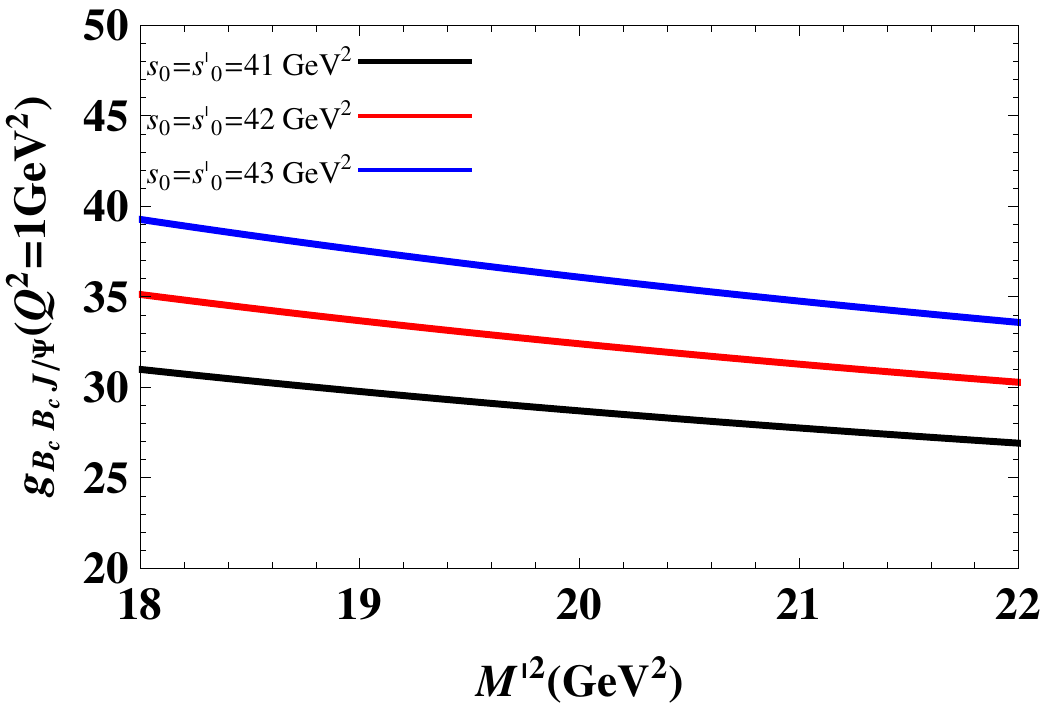}
\end{center}
\caption{The dependence of the form factor on $M'^{2}$.}
\label{fig3b}
\end{figure}

\begin{figure}[h!]
\begin{center}
\includegraphics[width=13cm]{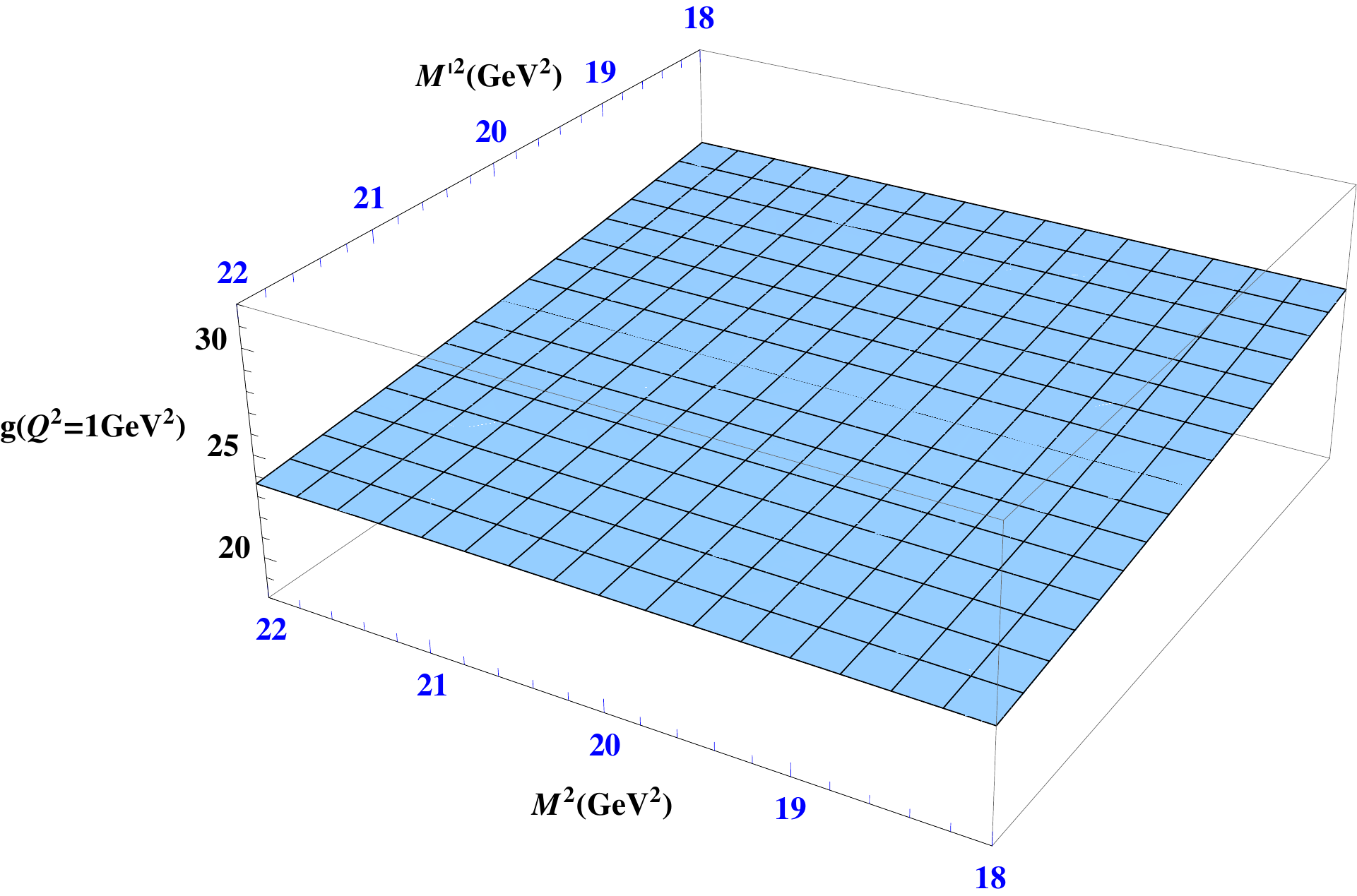}
\end{center}
\caption{The plateau of Borel mass parameters.}
\label{fig4}
\end{figure}

After determining the working regions of auxiliary parameters, the dependencies of strong form factor under consideration at $Q^2=-q^2=1GeV^2$ on Borel mass parameters $M^2$ and $M'^2$ are presented in Figs. (\ref{fig3a})-(\ref{fig4}).We see from these figures that the strong form factor is in good stability with respect to $M^2$ and $M'^2$ variations. By using other input parameters, we obtain the strong form factor is fitted to the following polynomial function:

\begin{eqnarray}\label{fitFunc}
g(Q^2)=\sum_i C_i Q^{2i}~ ,
\end{eqnarray}
for $i=0,1,2$ and $3$. The strong coupling constants are determined as the values of the form factors at $Q^2 = -m_{offshell}^2 = -9.3GeV^2$ at vacuum. Here, off-shell meson mass is the given mass of $m_{J/\psi}$ above. The obtained coupling constant in the framework of thermal QCD sum rules is $g_{B_c B_c J/\psi}=5.4\pm1.5$ which is consistent with previous theoretical calculations in the framework of sum rules at vacuum \cite{8}.

\begin{table}[h]
\renewcommand{\arraystretch}{1.5}
\addtolength{\arraycolsep}{3pt}
$$
\begin{array}{|c|c|c|c|}
\hline \hline
 C_0 &  C_1(GeV^{-2}) & C_2(GeV^{-4}) & C_3(GeV^{-6}) \\
\hline
  27.220       &  4.651   &  0.130  & -0.013 \\
 \hline \hline
\end{array}
$$
\caption{Fit function coefficients for the strong form factor. } \label{tab:mass}
\renewcommand{\arraystretch}{1}
\addtolength{\arraycolsep}{-1.0pt}
\end{table}

\begin{figure}[h!]
\begin{center}
\includegraphics[width=10cm]{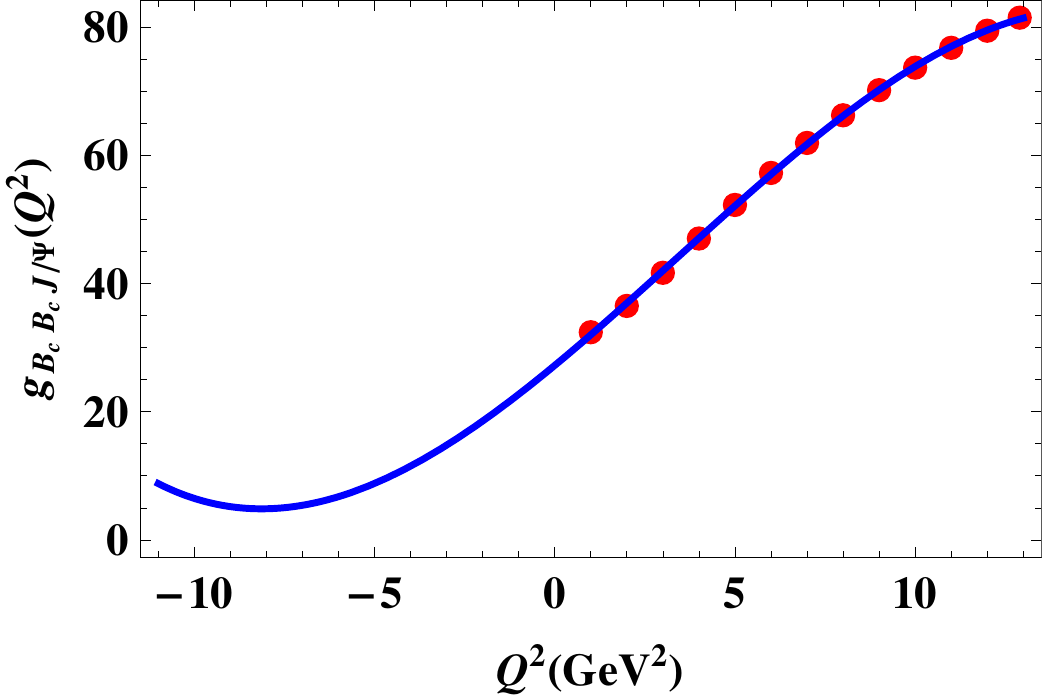}
\end{center}
\caption{The dependence of the form factor on $Q^2$ at $T=0~GeV$ (red dots) and the fitted curve.}
\label{gQsq}
\end{figure}

\begin{figure}[h!]
\begin{center}
\includegraphics[width=10cm]{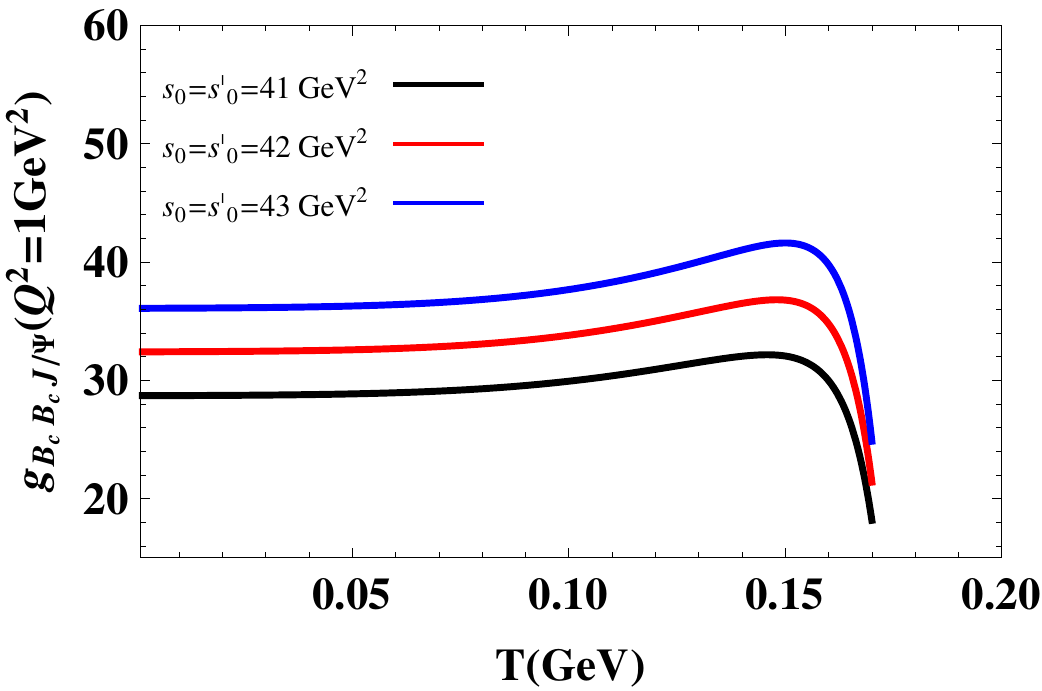}
\end{center}
\caption{The dependence of the form factor on temperature, for different continuum thresholds.}
\label{gTemp2}
\end{figure}

\begin{figure}[h!]
\begin{center}
\includegraphics[width=10cm]{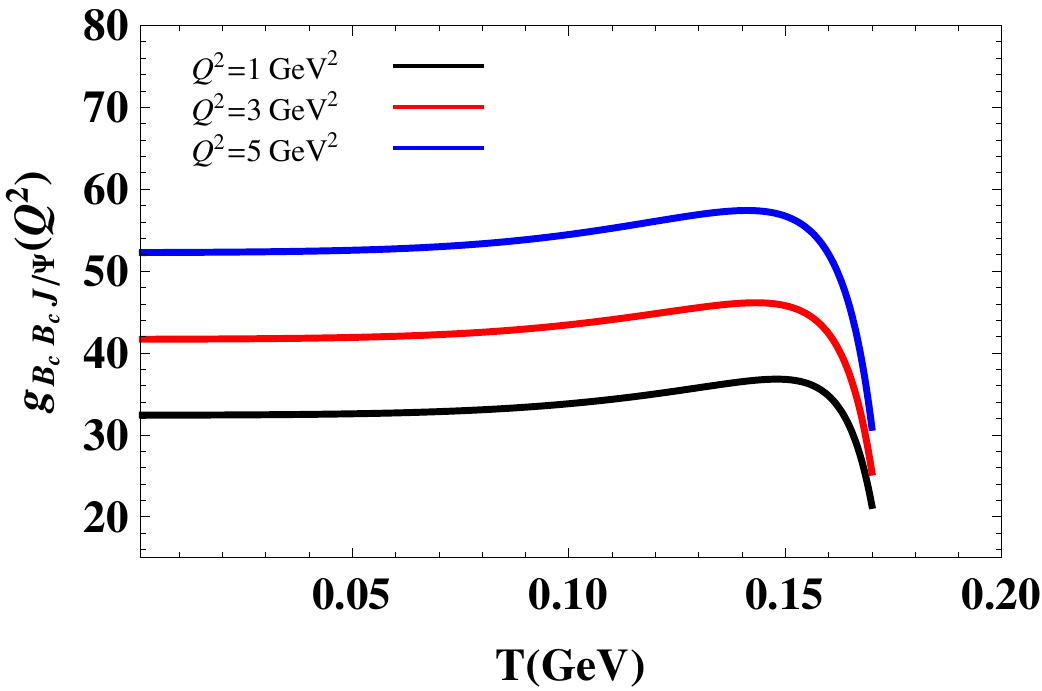}
\end{center}
\caption{The dependence of the form factor on temperature, for different $Q^2$ values ($s_0=42GeV^2$).}
\label{gTemp3}
\end{figure}

The dependence of the form factor of the vertex at vacuum is presented in Fig. (\ref{gQsq}). Our result for
the coupling constant at vacuum can be verified in the future experiments. The dependency of the form factor on the temperature is presented in Figs. (\ref{gTemp2}) and (\ref{gTemp3}) for different values of $Q^2$. The form factor does not change up to $T\simeq100~MeV$
 but slightly increases up to $T\simeq160~MeV$ then suddenly decreases  after this point.

\section{Conclusion}

The knowledge on temperature dependence of form factors is very important for interpretation of heavy-ion collision experiments and understanding QCD vacuum. In this article, we study the temperature dependence of the strong form factor of $B_cB_cJ/\psi$ vertex using the three-point QCD sum rules method. We also obtain the momentum dependence of the form factor at $T=0$, fit it into an analytic function and extrapolate into deep time-like region in order to obtain strong coupling constant of the vertex. The result is consistent with previous calculations \cite{8}. The coupling constant also can be used as a fundamental input parameter in researches of the heavy meson absorptions in hadronic matter. Besides, the behaviour of temperature dependency of the form factor can be checked in the future experiments.

\section{Acknowledgements}

This work has been supported in part by the Scientific and Technological Research Council of Turkey (TUBITAK) under the research projects 110T284 and 114F018.


%
%

%


%

\end{document}